\documentclass[prd,showpacs,twocolumn,superscriptaddress,nofootinbib,,floatfix,10pt]{revtex4-2}
\usepackage{setspace}
\usepackage[utf8]{inputenc}
\usepackage{float}
\usepackage{amsmath,amssymb,amsfonts,bm}
\usepackage{graphicx}
\usepackage{multirow}
\usepackage{slashed}
\usepackage{threeparttable}
\usepackage[usenames,dvipsnames]{color}
\allowdisplaybreaks
\usepackage[
colorlinks=true,
linkcolor=blue,
breaklinks=true,
urlcolor=blue,
citecolor=blue]{hyperref}

\usepackage{orcidlink}

\definecolor{green}{rgb}{0,0.6,0}

\newcommand{\ket}[1]{\left| #1 \right\rangle}

\newcommand{\be}{\begin{equation}} 
\newcommand{\ee}{\end{equation}}
\newcommand{\bea}{\begin{eqnarray}} 
\newcommand{\eea}{\end{eqnarray}}
\newcommand{\beas}{\begin{eqnarray*}} 
\newcommand{\eeas}{\end{eqnarray*}}

\renewcommand{\vec}{\bm}

\newcommand{\bonn}{\affiliation{Helmholtz-Institut f\"ur Strahlen- und Kernphysik and Bethe Center for Theoretical Physics,\\ Universit\"at Bonn, D-53115 Bonn, Germany}}

\newcommand{\fzj}{\affiliation{Institute for Advanced Simulation, Institut f\"ur Kernphysik and J\"ulich Center for Hadron Physics,\\ Forschungszentrum J\"ulich, D-52425 J\"ulich, Germany}}

\newcommand{\itp}{\affiliation{CAS Key Laboratory of Theoretical Physics, Institute of Theoretical Physics, \\Chinese Academy of Sciences, Beijing 100190, China}}

\newcommand{\ucas}{\affiliation{School of Physical Sciences, University of Chinese Academy of Sciences, Beijing 100049, China}}

\newcommand{\peng}{\affiliation{Peng Huanwu Collaborative Center for Research and Education, Beihang University, Beijing 100191, China}}

\newcommand{\Tbilisi}{\affiliation{Tbilisi State University, 0186 Tbilisi, Georgia}}

\newcommand{\scnt}{\affiliation{Southern Center for Nuclear-Science Theory (SCNT), Institute of Modern Physics,\\ Chinese Academy of Sciences, Huizhou 516000, China}}

\newcommand{\thu}{\affiliation{Department of Physics, Tsinghua University, Beijing 100084, China}}

\begin{document}
\author{Xiang-Kun Dong\orcidlink{0000-0001-6392-7143}}\email{xiangkun@hiskp.uni-bonn.de}\bonn

\author{Teng Ji\orcidlink{0000-0003-0366-1042}}\email{teng@hiskp.uni-bonn.de}\bonn

\author{Feng-Kun Guo\orcidlink{0000-0002-2919-2064}}\email{fkguo@itp.ac.cn}
\itp\ucas\peng \scnt

\author{Ulf-G. Mei{\ss}ner\orcidlink{0000-0003-1254-442X}}\email{meissner@hiskp.uni-bonn.de}
\bonn\fzj\Tbilisi

\author{Bing-Song~Zou\orcidlink{0000-0002-3000-7540}}\email{zoubs@itp.ac.cn}\itp \ucas \peng \scnt \thu 

\title{Hints of the $J^{PC}=0^{--}$ and $1^{--}$ $K^*\bar K_1(1270)$ Molecules in the $J/\psi\to\phi\eta\eta'$ Decay}

\begin{abstract}
The primary objective of this study is to investigate  hadronic molecules of $K^*\bar K_1(1270)$ using a one-boson-exchange model, which incorporates exchanges of vector and pseudoscalar mesons in the $t$-channel, as well as the pion exchange in the $u$-channel. Additionally, careful consideration is given to the three-body effects resulting from the on-shell pion originating from $K_1(1270)\to K^*\pi$. Then the BESIII data of the $J/\psi\to\phi\eta\eta'$ process is fitted using the  $K^*\bar K_1(1270)$ scattering amplitude with $J^{PC}=0^{--}$ or $1^{--}$. The analysis reveals that both the $J^{PC}=0^{--}$ and $1^{--}$ assumptions for $K^*\bar K_1(1270)$ scattering provide good descriptions of the data, with similar fit qualities. Notably, the parameters obtained from the best fits indicate the existence of $K^*\bar K_1(1270)$ bound states, denoted by $\phi(2100)$ and $\phi_0(2100)$ for the $1^{--}$ and $0^{--}$ states, respectively. The current experimental data, including the $\eta$ polar angular distribution, cannot distinguish which $K^*\bar K_1(1270)$ bound state contributes to the $J/\psi\to\phi\eta\eta'$ process, or if both are involved. Therefore, we propose further explorations of this process, as well as other processes, in upcoming experiments with many more $J/\psi$ events to disentangle the different possibilities.
\end{abstract}

\maketitle

\section{Introduction}
Exotic hadrons, which lie beyond the conventional quark model~\cite{GellMann:1964nj,Zweig:1964jf}, have gained significant attention in the past two decades due to the observation of numerous exotic states or their candidates in experiments. Despite of extensive research on the structures and properties of these exotic states, many of them remain subjects of debate. We refer to Refs.~\cite{Chen:2016qju,Hosaka:2016pey,Richard:2016eis,Lebed:2016hpi,Esposito:2016noz,Guo:2017jvc,Ali:2017jda,Olsen:2017bmm,Belle-II:2018jsg,Cerri:2018ypt,Liu:2019zoy,Brambilla:2019esw,Guo:2019twa,Yang:2020atz,Dong:2021juy,Chen:2022asf,Dong:2021bvy,Yamaguchi:2019vea} for recent reviews on the experimental and theoretical status of exotic hadrons. One intriguing observation is that many of the observed peaks are located very close to the thresholds of hadron pairs that they can couple to. This proximity can be attributed to the $S$-wave attraction between the relevant hadron pair, as discussed in Ref.~\cite{Dong:2020hxe}. Consequently, a natural interpretation for these states is the formation of hadronic molecules, as extensively reviewed in Refs.~\cite{Chen:2016qju,Guo:2017jvc,Brambilla:2019esw,Yamaguchi:2019vea,Dong:2021juy,Dong:2021bvy}. 

Among the exotic states, those with exotic quantum numbers $J^{PC}$ that cannot be formed by conventional quark-antiquark mesons, such as $0^{--},1^{-+}$ and so on, are of extremely great interest. Currently, there have been four experimental candidates of such exotic states, namely $\pi_1(1400)$, $\pi_1(1600)$~\cite{Meyer:2015eta}, $\eta_1(1855)$~\cite{BESIII:2022riz} and $\pi_1(2015)$~\cite{E852:2004gpn,E852:2004rfa}, all possessing $J^{PC}=1^{-+}$. Although numerous theoretical studies have proposed the existence of $0^{--}$ states, such as compact tetraquark states~\cite{Cotanch:2006wv,General:2007bk,Maiani:2014aja,Cleven:2015era,Wang:2021lkg}, hybrid states~\cite{Ishida:1991mx,General:2006ed,Govaerts:1984bk,Chetyrkin:2000tj,Huang:2016rro,Liu:2005rc}, glueballs~\cite{Qiao:2014vva,Pimikov:2016pag,Pimikov:2017bkk,Pimikov:2017xap}, or a $D^*\bar D^*_0$ hadronic molecule~\cite{Shen:2010ky}, no experimental signals have been reported thus far. One should notice that the above predictions may have large uncertainties and some of them are still controversial, even problematic. For example, the QCD sum rules concluded that no $0^{--}$ tetraquark state exists below 2~GeV~\cite{LEE:2020eif,Jiao:2009ra}. 

In a recent work~\cite{Ji:2022blw}, a narrow $0^{--}$ $D^*\bar D_1(2420)$ molecule $\psi_0(4360)$ was predicted in the one-boson-exchange (OBE) model based on heavy quark spin symmetry and the assumption that the $\psi(4230)$, $\psi(4360)$ and $\psi(4415)$ can be identified as hadronic molecules consisting of $D\bar D_1$, $D^*\bar D_1(2420)$ and $D^*\bar D^*_2$ components, respectively~\cite{Wang:2013cya,Wang:2013kra,Ma:2014zva,Cleven:2015era,Hanhart:2019isz,Anwar:2021dmg}. This predicted state can be searched for in the $J/\psi\eta$ or $D\bar D^*$ final states in $e^+e^-$ collisions, specifically in the processes $e^+e^-\to J/\psi\eta\eta$ or $e^+e^-\to D\bar D^*\eta$. Analogously, in the hidden strangeness channel, we can investigate the $0^{--}$ molecule composed of $K^*\bar K_1$. These states may manifest in the $\phi\eta^{(\prime)}$ or $K\bar K^*$ final states in the decays of $J/\psi$.

In Ref.~\cite{BESIII:2018zbm}, a total of $1.3\times10^9$ $J/\psi$ events were used to investigate the decay process $J/\psi\to\phi\eta'\eta$. Notably, an enhancement around 2.1~GeV was observed in the final states involving $\phi\eta'$. By incorporating a Breit-Wigner (BW) resonance with $J^{P}=1^{+}$ or $1^-$, the invariant mass distribution of $\phi\eta'$ was well described, while the possibility of $J^{P}=0^{-}$ was ruled out based on the distribution of the $\eta$ polar angle, which represents the angle between the outgoing $\eta$ meson and the incoming $e^+e^-$ beams in the rest frame of the $J/\psi$. However, we will later explain that the current data do not provide conclusive evidence to exclude the $J^P=0^-$ possibility due to the significant contribution of the phase space (PHSP) processes derived from the experimental Monte Carlo simulations.

In the Review of Particle Physics (RPP)~\cite{ParticleDataGroup:2022pth}, there are two $K_1$ particles, namely $K_1(1270)$ and $K_1(1400)$. Given that the observed enhancement in $J/\psi\to\phi\eta'\eta$ is slightly below the threshold of $K^*\bar K_1(1270)$, it is reasonable to investigate whether the $\phi\eta'$ invariant mass distribution can be explained by the presence of $K^*\bar K_1(1270)$ molecular states. In the following analysis, we will use $K_1$ to refer to $K_1(1270)$ unless otherwise specified.

\section{$K^*\bar K_1$ scattering in the OBE model}\label{sec:OBE}

\subsection{The $K^*\bar K_1$ potentials}\label{sec:potential}
The flavor wave function of the $K^*\bar K_1 $ state with specific $J^{PC}$ can be expressed as
\begin{align}
    \ket{K^*K_1}_{J^{PC}}&=\frac{1}{\sqrt2}\left(\ket{K^*\bar K_1}+C(-1)^{J-J_1-J_2}\mathcal C\ket{K^* \bar K_1}\right),
\end{align}
where $J_1$ represents the spin of $K^*$, $J_2$ represents the spin of $K_1$, and $\mathcal C$ refers to the charge conjugation operator. 
Using the following phase conventions for the charge conjugation transformation,
\begin{align}
    \mathcal{C}\ket{K^*}=-\ket{\bar{K^*}},\quad \mathcal{C}\ket{K_1}=\ket{\bar{K_1}}\label{eq:Ccon},
\end{align}
we have
\begin{align}
    \ket{K^*K_1}_{1^{--}}&=\frac{1}{\sqrt2}\left(\ket{K^*\bar K_1}+\ket{\bar K^* K_1}\right),\\
    \ket{K^*K_1}_{0^{--}}&=\frac{1}{\sqrt2}\left(\ket{K^*\bar K_1}-\ket{\bar K^* K_1}\right).
\end{align}

In order to assess the exchanges of the vector meson ($V$) and pseudoscalar meson ($P$)  between $K^*$ and $\bar K_1$ in the $t$-channel, the Lagrangian of $K^*K^*V/P$ coupling is needed. From the hidden local symmetry formalism, the relevant Lagrangians can be constructed as~\cite{Meissner:1987ge,Bando:1987br,Bando:1984ej}
\begin{align}
\mathcal{L}_{V V V}&=\,i g\langle\left(\partial_\mu V_\nu-\partial_\nu V_\mu\right) V^\mu V^\nu\rangle\label{eq:Lagran_VVV},\\
\mathcal{L}_{V V P}&=\frac{G^{\prime}}{\sqrt{2}} \epsilon^{\mu \nu \alpha \beta}\left\langle\partial_\mu V_\nu \partial_\alpha V_\beta P\right\rangle \label{eq:Lagran_VVP},
\end{align}
where
\begin{align}
    V^\mu&=\left(\begin{array}{ccc}
\frac{1}{\sqrt{2}} \rho^0+\frac{1}{\sqrt{2}} \omega & \rho^{+} & K^{*+} \\
\rho^{-} & -\frac{1}{\sqrt{2}} \rho^0+\frac{1}{\sqrt{2}} \omega & K^{* 0} \\
K^{*-} & \bar{K}^{* 0} & \phi
\end{array}\right)^\mu,\\
P&=\left(\begin{array}{ccc}
\frac{1}{\sqrt{2}} \pi^0+\frac{1}{\sqrt{6}} \eta & \pi^{+} & K^{+} \\
\pi^{-} & -\frac{1}{\sqrt{2}} \pi^0+\frac{1}{\sqrt{6}} \eta & K^0 \\
K^{-} & \bar{K}^0 & -\frac{2}{\sqrt{6}} \eta
\end{array}\right),
\end{align}
and $\langle\cdots\rangle$ means the trace in flavor space. The coupling constant $g$ is expressed as $g={m_V}/({2F_\pi})$ where $m_V$ represents the mass of the vector meson $\rho$ and $F_\pi=92.4$ MeV is the pion decay constant. The coupling constant $G'$ is expressed as $G'={3g^{\prime 2}}/({4\pi^2F_\pi})$ with $g'=-{G_Vm_V}/({\sqrt2F_\pi^2})$ and $G_V={F_\pi}/{\sqrt{2}}$~\cite{Ecker:1989yg}.

Expanding Eqs.~\eqref{eq:Lagran_VVV} and~\eqref{eq:Lagran_VVP}, we  obtain the following $K^*K^*V/P$ couplings,
\begin{align}
\mathcal{L}_{ K^* K^*\rho} &=\frac{ig}{\sqrt2}\big\{\left[\bar{K}^*_{\nu} \vec{\tau}\left(\partial_\mu K^{*\nu}\right)-\left(\partial_\mu \bar{K}^*_{\nu}\right) \vec{\tau} K^{*\nu}\right] \cdot \vec{\rho}^\mu\notag \\
&\quad\quad\quad\ +2\bar{K}^*_{\nu} \vec{\tau}K^*_{\mu}\cdot\left(\partial^\nu\bm{\rho}^\mu-\partial^\mu\bm{\rho}^\nu\right)\big\},\\
\mathcal{L}_{ K^* K^*\omega}&=\frac{ig}{\sqrt2}\big\{\left[\bar{K}^*_{\nu}\left(\partial_\mu K^{*\nu}\right)-\left(\partial_\mu \bar{K}^*_{\nu}\right) K^{*\nu}\right] \omega^\mu\notag\\
&\quad\quad\quad\ +2\bar{K}^*_{\nu}K^*_{\mu}\left(\partial^\nu{\omega}^\mu-\partial^\mu{\omega}^\nu\right)\big\},\\
\mathcal{L}_{ K^* K^*\phi}&=- ig \big\{\left[\bar{K}^*_{\nu}\left(\partial_\mu K^{*\nu}\right)-\left(\partial_\mu \bar{K}^*_{\nu}\right) K^{*\nu}\right] \phi^\mu\notag\\
&\quad\quad\quad\ +2\bar{K}^*_{\nu}K^*_{\mu}\left(\partial^\nu{\phi}^\mu-\partial^\mu{\phi}^\nu\right)\big\},\\
\mathcal{L}_{ K^* K^*\pi} &=\frac{G^{\prime}}{2} \epsilon^{\mu \nu \alpha \beta}\partial_\mu \bar{K}^*_{\nu}{\bm\tau\cdot\bm\pi}\partial_\alpha K^*_{\beta},\\
\mathcal{L}_{ K^* K^*\eta}&= -\frac{G^{\prime}}{2\sqrt{3}} \epsilon^{\mu \nu \alpha \beta}\partial_\mu \bar{K}^{*}_\nu\partial_\alpha K^*_\beta\eta,
\end{align}
with
\begin{align}
K^*&=\left(\begin{array}{c}{K^{*+}} \\ {K^{*0}}\end{array}\right), \bar{K}^*=\left(K^{*-}, \bar{K}^{*0}\right),\\ \vec{\rho}&=\left(\frac{\rho^++\rho^-}{\sqrt{2}},\frac{\rho^--\rho^+}{i\sqrt{2}},\rho^0\right),\\
\vec{\pi}&=\left(\frac{\pi^++\pi^-}{\sqrt{2}},\frac{\pi^--\pi^+}{i\sqrt{2}},\pi^0\right),
\end{align}
and $\vec{\tau}$ are the Pauli matrices in the isospin space. 

We assume that the $K_1K_1V/P$ couplings have the same form as the $K^*K^*V/P$ couplings,
\begin{align}
\mathcal{L}_{ K_1 K_1\rho} &=\frac{ig_1}{\sqrt2}\big\{\left[\bar{K}_{1\nu} \vec{\tau}\left(\partial_\mu K^{\nu}_1\right)-\left(\partial_\mu \bar{K}_{1\nu}\right) \vec{\tau} K^{\nu}_1\right] \cdot \vec{\rho}^\mu\notag \\
&\quad\quad\quad\ +2\bar{K}_{1\nu} \vec{\tau}K_{1\mu}\cdot\left(\partial^\nu\bm{\rho}^\mu-\partial^\mu\bm{\rho}^\nu\right)\big\},\\
\mathcal{L}_{ K_1 K_1\omega}&=\frac{ig_1}{\sqrt2}\big\{\left[\bar{K}_{1\nu}\left(\partial_\mu K^{\nu}_1\right)-\left(\partial_\mu \bar{K}_{1\nu}\right) K^{\nu}_1\right] \omega^\mu\notag\\
&\quad\quad\quad\ +2\bar{K}_{1\nu}K_{1\mu}\left(\partial^\nu{\omega}^\mu-\partial^\mu{\omega}^\nu\right)\big\},\\
\mathcal{L}_{ K_1 K_1\phi}&=- ig_1 \big\{\left[\bar{K}_{1\nu}\left(\partial_\mu K^{\nu}_1\right)-\left(\partial_\mu \bar{K}_{1\nu}\right) K^{\nu}_1\right] \phi^\mu\notag\\
&\quad\quad\quad\ +2\bar{K}_{1\nu}K_{1\mu}\left(\partial^\nu{\phi}^\mu-\partial^\mu{\phi}^\nu\right)\big\},\\
\mathcal{L}_{ K_1 K_1\pi} &=\frac{G_1^{\prime}}{2} \epsilon^{\mu \nu \alpha \beta}\partial_\mu \bar{K}_{1\nu}{\bm\tau\cdot\bm\pi}\partial_\alpha K_{1\beta},\\
\mathcal{L}_{ K_1 K_1\eta}&= -\frac{G_1^{\prime}}{2\sqrt{3}} \epsilon^{\mu \nu \alpha \beta}\partial_\mu \bar{K}_{1\nu}\partial_\alpha K_{1\beta}\eta,
\end{align}
with
\begin{align}
K_1&=\left(\begin{array}{c}{K^{+}_1} \\ {K^{0}_1}\end{array}\right), \bar{K}_1=\left(K^{-}_1, \bar{K}^{0}_1\right).
\end{align}
We further assume 
that the coupling constants $g_1$ and $G'_1$ should be of the same order as $g$ and $G'$, respectively. As a result, we opt to set $g_1=g$ and $G'_1=G'$ in the following calculations. This would be the case in the massive Yang-Mills model for vector mesons \cite{Meissner:1987ge}. We have  verified that any deviation of approximately 20\% in $g_1$ and $G'_1$ can be adequately accommodated by varying the cutoff to be introduced later.

Using the above Lagrangian, we obtain the potentials from $t$-channel meson exchanges in momentum space,
\begin{widetext}
\begin{align}
    V_{V,t}^{(1)}(\bm q)&=-\frac{g^2}{2}\left(1-\frac{\bm q^2}{3}\left(\frac{1}{\mu^2}-\frac{4}{m_{K^{*}} m_{K_1}}\right)\right)\frac{F_V^t}{\bm q^2+m_V^2+i\epsilon} = A_V^{(1)}-\frac{g^2}{2}\left(1+\frac{m_V^2}{3}\left(\frac{1}{\mu^2}-\frac{4}{m_{K^{*}} m_{K_1}}\right)\right)\frac{F_V^t}{\bm q^2+m_V^2+i\epsilon}\label{eq:VVt1},\\
    V_{V,t}^{(0)}(\bm q)&=-\frac{g^2}{2}\left(1-\frac{\bm q^2}{3}\left(\frac{1}{\mu^2}-\frac{6}{m_{K^{*}} m_{K_1}}\right)\right)\frac{F_V^t}{\bm q^2+m_V^2+i\epsilon} = A_V^{(0)}-\frac{g^2}{2}\left(1+\frac{m_V^2}{3}\left(\frac{1}{\mu^2}-\frac{6}{m_{K^{*}} m_{K_1}}\right)\right)\frac{F_V^t}{\bm q^2+m_V^2+i\epsilon}\label{eq:VVt0},\\
    V_{P,t}^{(1)}(\bm q)&=\frac{G^{'2}}{48}\frac{F_P^t \bm q^2}{\bm q^2+m_P^2+i\epsilon} = -\frac{G^{'2}}{48}\frac{F_P^t m_P^2}{\bm q^2+m_P^2+i\epsilon}+A_P^{(1)}\label{eq:VPt1},\\
    V_{P,t}^{(0)}(\bm q)&=2 \  V_{P,t}^{(1)}(\bm q)=-\frac{G^{'2}}{24}\frac{F_P^t m_P^2}{\bm q^2+m_P^2+i\epsilon}+A_P^{(0)}, \label{eq:VPt0}
\end{align}
\end{widetext}
where
\begin{align}
    A_V^{(1)}&=\frac{g^2 F_V^t}{2}\left(\frac{1}{\mu^2}-\frac{4}{m_{K^{*}} m_{K_1}}\right),\\
    A_V^{(0)}&=\frac{g^2 F_V^t}{2}\left(\frac{1}{\mu^2}-\frac{6}{m_{K^{*}} m_{K_1}}\right),\\
   A_P^{(0)}&= 2A_P^{(1)}=\frac{G^{'2}}{24} F_P^t,
\end{align}
and $\mu$ is the reduced mass of $K^*\bar K_1$ and $\bm q=\bm k-\bm k'$ is the three-momentum of the exchanged $\pi$ with $\bm k$ and $\bm k'$ the three-momenta of the incoming and outgoing particles in the center-of-mass (c.m.) frame, respectively. The superscripts $(1)$ and $(0)$ represent the results in the $1^{--}$ and $0^{--}$ cases, respectively. The flavor factors are $F_\rho^t=3,F_\omega^t=1,F_\phi^t=2,F_\pi^t=3$ and $F_\eta^t=1/3$. The constant components of the potentials, $A_{V/P}^{(1/0)}$, will be rewritten as two scale-dependent parameters~\cite{Ji:2022blw,Yalikun:2021bfm},
\begin{align}
    C^{(1)}(\Lambda)=c(\Lambda)\sum_{V={\rho,\omega,\phi}}A_V^{(1)}+d(\Lambda)\sum_{P={\pi,\eta}}A_P^{(1)},\label{eq:C1}\\
    C^{(0)}(\Lambda)=c(\Lambda)\sum_{V={\rho,\omega,\phi}}A_V^{(0)}+d(\Lambda)\sum_{P={\pi,\eta}}A_P^{(1)},\label{eq:C2}
\end{align}
which will serve as counterterms to absorb the cutoff $(\Lambda)$ dependence as will be explained later. We will take $C^{(1)}$ and  $C^{(0)}$ as free parameters to be fitted. 

The $S$-wave $K_1K^*\pi$ coupling can be expressed as\footnote{In principle, there should be other terms containing $\partial_\mu \partial_\nu \pi$, which are, however, one order higher than those in Eq.~\eqref{eq:Vpiu} in the power expansion of the pion momentum. Therefore, these terms are omitted.}
\begin{align}
    \mathcal L_{K_1K^*\pi}=i\frac{g_S}{\sqrt{2}}\left(\partial_\nu \bar K_{1}^\mu\bm\tau K^*_\mu- \bar K_{1}^\mu\bm\tau \partial_\nu K^*_\mu\right)\cdot\partial^\nu\bm\pi+\rm h.c.,\label{eq:Vpiu}
\end{align}
where the coupling constant $g_S=3.4$ is determined by the partial decay width of $K_1 \to K^*\pi$. Note that we have ignored a possible $D$-wave contribution. Utilizing the Lagrangian in Eq.~\eqref{eq:Vpiu}, we obtain the potential for the $u$-channel $\pi$ exchange as
\begin{align}
    V_{\pi,u}^{(1)}(q)&=\frac{3}{8}g_S^2\frac{(m_{K_1}^2-m_{K^*}^2)^2}{4m_{K_1}m_{K^*}}\frac{1}{q^2-m_\pi^2+i\epsilon},\\
    V_{\pi,u}^{(0)}(q)&=-V_{\pi,u}^{(1)}(q),
\end{align}
where $q$ represents the four-momentum of the exchanged pion.

\subsection{Lippmann-Schwinger Equation}
The scattering amplitude can be obtained by solving the Lippmann-Schwinger Equation (LSE),
\begin{align}
T(E ; \boldsymbol{k}^{\prime}, \boldsymbol{k})&=V\left(E;\boldsymbol{k}^{\prime}, \boldsymbol{k}\right)\notag\\
&+\int \frac{{\rm d}^{3} \boldsymbol{l}}{(2 \pi)^{3}} \frac{V\left(E ;\boldsymbol{k}^{\prime}, \boldsymbol{l}\right) T(E ; \boldsymbol{l}, \boldsymbol{k})}{E-\boldsymbol{l}^{2} /\left(2 \mu\right)+i \Gamma(E;\boldsymbol{l})/2},\label{eq:LSE}
\end{align}
where $\bm k$ and $\bm{k'}$ are the three-momenta of the initial and final states in the c.m. frame, in order, $\mu$ is the reduced mass of $K^*\bar K_1$, and $E$ is the energy relative to the threshold. The energy-dependent width $\Gamma(E;\boldsymbol{l})$ is the sum of the widths of $K^*$ and $K_1$. The integral is ultraviolet divergent and it is regularized by introducing  a Gaussian form factor,
\begin{align}
    V\left(E;\boldsymbol{k}^{\prime}, \boldsymbol{k}\right)\to V\left(E;\boldsymbol{k}^{\prime}, \boldsymbol{k}\right)e^{-\bm{q}^2/\Lambda^2},\label{eq:Vcut}
\end{align}
where $\Lambda$ is the cutoff parameter. The effects of the variation of $\Lambda$ can be absorbed by adjusting the value of $C^{(1)}$ or $C^{(0)}$ introduced in Eqs.~(\ref{eq:C1}, \ref{eq:C2}).

After the  $S$-wave projection, the LSE in Eq.~\eqref{eq:LSE} is reduced to 
\begin{align}
    T_0(E ; {k}^{\prime},  {k})=&\,V_0\left(E; {k}^{\prime},  {k}\right)\notag\\
&+\int \frac{{\rm d}  {l}}{2 \pi^{2}} \frac{l^2V_0\left(E ; {k}^{\prime},  {l}\right) T_0(E ;  {l},  {k})}{E- {l}^{2} /\left(2 \mu\right)+i \Gamma(E; {l})/2},\label{eq:LSE0}
\end{align}
with $k, k'$ and $l$ the magnitudes of the corresponding three-momenta.
We would like to emphasize that the $S$-wave projection of the potential is nontrivial, and it will result in additional cuts to the scattering amplitude~\cite{Du:2021zzh,Ji:2022blw}. This introduces significant complexity, particularly for the $u$-channel $\pi$ exchange. Further details can be found in Appendix~\ref{app:SwavePro}.

The $K^*$ dominantly decays into $K\pi$ with a decay width around $\Gamma_{K^*}=50$~MeV~\cite{ParticleDataGroup:2022pth}. The total decay width of the $K_1$ is $(90\pm20)$~MeV and the branching ratio of $K_1\to K^*\pi$ is $(21\pm 10)\%$~\cite{ParticleDataGroup:2022pth}.\footnote{The central values are used in the following calculations.}  For simplicity, in $\Gamma(E; {l})$ we only include the energy dependence of the partial width of $K_1\to K^*\pi$ since this process contribute to the pion exchange between $K^*$ and $\bar K_1$ in the $u$-channel. Explicitly, we have
\begin{align}
\Gamma(E; {l})&=\Gamma_{K^*}+\Gamma_{K_1}(E; {l}),\\
    \Gamma_{K_1}(E; {l})&=g_S^2\frac{(m_{K^*\pi}^2-m_{K^*}^2)^2}{8}\frac{q_{\rm eff}(E; {l})}{8\pi m_{K^*\pi}^2}+\Gamma_{K_1}^{\rm cons},
\end{align}
where $\Gamma_{K_1}^{\rm cons}=71$~MeV is the decay width of $K_1$ apart from the $K^*\pi$ channel, 
\begin{align}
    m_{K^*\pi}=E+m_{K_1}- \frac{l^2}{2\mu}
\end{align}
is the invariant mass of $K^*\pi$ from the $K_1$ decay, and $q_{\rm eff}(E; {l})$ is the momentum of the $\pi$ in the rest frame of $K_1$, determined by
\begin{align}
    q_{\rm eff}(E; {l})&=q_{\rm cm}(m_{K^*\pi}(E; {l}),m_{K^*},
    m_{\pi}).
\end{align}
The function
\begin{align}
    q_{\rm cm}(M,m_1,m_2)=\frac{1}{2M}\sqrt{\lambda(M^2,m_1^2,m_2^2)}\label{eq:qcm}
\end{align}
yields the momentum of $m_1$ in the rest frame of $M$ in the decay process of $M\to m_1m_2$ and $\lambda(x,y,z)=x^2+y^2+z^2-2xy-2yz-2zx$ is the K\"all\'en triangle function. The $K^*\pi$ loop in the $K_1$ propagator introduces an additional cut, which is represented by a three-body cut extending from the $K^*\bar K^*\pi$ threshold to infinity. To ensure a smooth crossing of this cut when searching for poles in the complex energy plane, the cut of the square root function in Eq.~\eqref{eq:qcm} is defined along the negative imaginary axis~\cite{Doring:2009yv,Ji:2022blw,Du:2021zzh}.

\section{Fitting the $J/\psi\to\phi\eta'\eta$ data}\label{sec:fitting}
\subsection{$J/\psi\to\phi\eta'\eta$ amplitude with $K^*\bar K_1$ rescattering}
\begin{figure}
    \centering    \includegraphics[width=\linewidth]{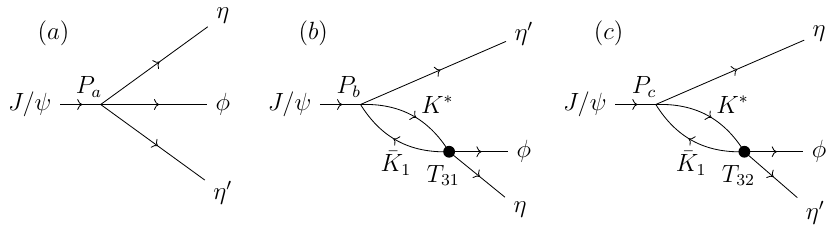}
    \caption{Diagrams of the $J/\psi\to\phi\eta'\eta$ decay with intermediate $K^*\bar K_1$ rescattering. }
    \label{fig:feyndiag}
\end{figure}

The $S$-wave $K^*\bar K_1$ system can couple to both the $\phi\eta'$ and $\phi\eta$ final states. In the following analysis, we consider the interaction between the $\phi\eta$, $\phi\eta'$ and $K^*\bar K_1$ coupled channels, which are labeled as channel 1, 2 and 3, respectively. The scattering amplitudes are described by the coupled-channel LSE,
\begin{align}
    T_{ij}=V_{ij}+\sum_{k=1}^{3}V_{ik}G_{kk}T_{kj},
\end{align}
where $G_{kk}$ represents the loop function of the two-particle propagators of channel $k$. The term $V_{33}$ is the $K^*\bar K_1$ potential obtained in Sect.~\ref{sec:potential}.  Since the $J^{PC}$ of the system is either $1^{--}$ or $0^{--}$, the $\phi\eta^{(\prime)}$ must be in $P$-wave. Therefore we neglect the interaction between $\phi\eta^{(\prime)}$ which is not expected to alter the existence of the $K^*\bar K_1$ molecular states. Similarly, we expect $V_{31}$ and $V_{32}$ to be small and treat them in perturbation theory. Consequently, the potential matrix reads
\begin{align}
    V=\left(\begin{array}{ccc}
        0 & 0 & v_{31}\tilde q_{\eta} \\
        0 & 0 & v_{32}\tilde q_{\eta'}\\
        v_{31}\tilde q_{\eta} & v_{32}\tilde q_{\eta'} & V_{33}
    \end{array}\right),
\end{align}
where $v_{31}$ and $v_{32}$ are constants, $\tilde q_{\eta^{(\prime)}}$ represents the three-momentum of $\eta^{(\prime)}$ in the $\phi\eta^{(\prime)}$ c.m. frame. We thus have
\begin{align}
T_{33}&=V_{33}+V_{33}G_{33}T_{33}+\mathcal{O}\left(V_{31}^2,V_{32}^2\right),\\
T_{31}&=T_{33}V_{33}^{-1}V_{31}+\mathcal{O}\left(V_{31}^3,V_{32}^3\right),\\
T_{32}&=T_{33}V_{33}^{-1}V_{32}+\mathcal{O}\left(V_{31}^3,V_{32}^3\right).
\end{align}
Upon disregarding the $\mathcal{O}\left(V_{31}^2,V_{32}^2\right)$ terms, it becomes apparent that $T_{33}$ corresponds to the single-channel $K^*\bar K_1$ scattering amplitude, which has been derived in Eq.~\eqref{eq:LSE0}. The process of $K^*\bar K_1\to\phi\eta^{(\prime)}$ inelastic scattering can be approximated as $T_{33}V_{33}^{-1}V_{31}$ or $T_{33}V_{33}^{-1}V_{32}$. Here, both $T_{33}$ and $V_{33}$ are known, and the constants $v_{31}$ and $v_{32}$ can be absorbed into the normalization constant of the experimental data during the fitting process.

From the results obtained in Ref.~\cite{BESIII:2018zbm}, the contribution of the $f_0(1500)$ in $\eta\eta'$ can be considered negligible. Consequently, the amplitude of $J/\psi\to\phi\eta'\eta$ can be represented as follows,
\begin{align}
T_{J/\psi\to\phi\eta'\eta}=P_{a}q_{\eta}\tilde q_{\eta'}+P_bG_{33}T_{31}q_{\eta'}+P_cG_{33}T_{32}q_{\eta},\label{eq:Tjpsi}
\end{align}
where $q_{\eta'}$ denotes the three-momentum of the $\eta'$ in the $J/\psi$ rest frame in Fig.~\ref{fig:feyndiag}$(b)$, whereas $q_{\eta}$ denotes the three-momentum of the $\eta$ in the $J/\psi$ rest frame in Fig.~\ref{fig:feyndiag}$(c)$. $P_{a}$, $P_{b}$ and $P_c$ are constants that represent the production parameters of $\phi\eta'\eta$, $K^*\bar K_1\eta'$ and $K^*\bar K_1\eta$ in the decay of $J/\psi$, respectively. Note that we introduce the additional momentum $q_{\eta^{(')}}$ due to the fact that the $\eta'$ in Fig.~\ref{fig:feyndiag}$(b)$ is in $P$-wave, so is the $\eta$ in Fig.~\ref{fig:feyndiag}$(c)$. The $P_a$ term represents the production of the $P$-wave $\eta$ and $\phi\eta'$, which is in fact a higher order term. The leading contribution from the $S$-wave $\phi\eta'\eta$ production leads to a constant contact term and is covered by the background to be introduced in Eq.~\eqref{eq:dGamma}.  The loop propagator of the $K^*\bar K_1$ channel reads
\begin{align}
    G_{33}=\int\frac{{\rm d}\bm l^3}{(2\pi)^3}\frac{e^{-\bm l^2/\Lambda_1^2}}{E-\bm l^2/(2\mu)+i\Gamma(E;\bm l)/2},
\end{align}
where the cutoff $\Lambda_1$ is fixed to $1$~GeV. The $\Lambda_1$-dependence of the physical results will be absorbed by the production parameters.

The differential decay width of $J/\psi$ is now expressed as
\begin{align}
\frac{\mathrm{d}\Gamma_{J/\psi\to\phi\eta'\eta}}{\mathrm{d}M_{\phi\eta'}}&=\int{\rm d}M_{\phi\eta}^2\frac{2M_{\phi\eta'}}{256\pi^3m^3_{J/\psi}}|T_{J/\psi\to\phi\eta'\eta}|^2\notag\\
&\ \ \ \ +\alpha f_{\rm bg}(M_{\phi\eta'}).\label{eq:dGamma}
\end{align}
We have introduced a noninterfering background term $\alpha f_{\rm bg}(M_{\phi\eta'})$, where $f_{\rm bg}(M_{\phi\eta'})$ mimics the lineshape of the PHSP process determined by the Monte Carlo simulation in Ref.~\cite{BESIII:2018zbm}. The parameter $\alpha$ represents the magnitude that needs to be fitted.
It can come from a purely $S$-wave production term, which does not interfere with the $P$-wave ones in Eq.~\eqref{eq:Tjpsi}.

\subsection{$K^*\bar K_1$ in only $\phi\eta'$ channel}
\begin{figure*}[t]
    \centering
    \includegraphics[width=0.8\linewidth]{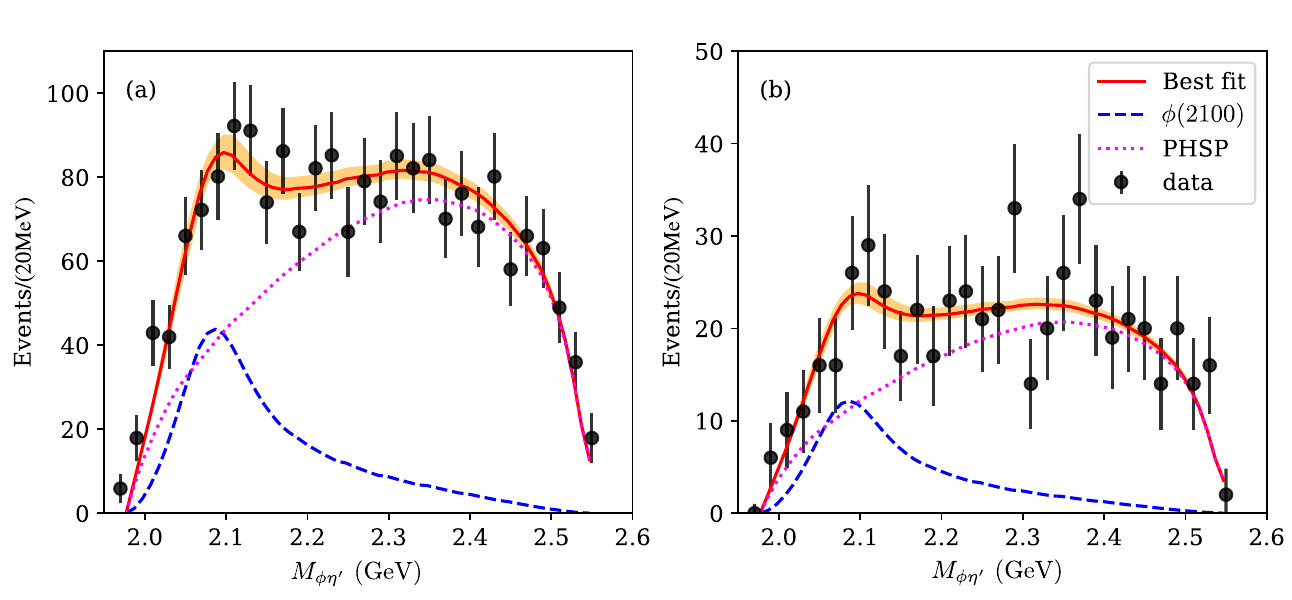}
    \caption{The best fit of the $J/\psi\to\phi\eta'\eta$ data~\cite{BESIII:2018zbm} with $1^{--}$ $K^*\bar K_1$ rescattering only in the $\phi\eta'$ channel ($P_b=0$). The $\eta'$ is reconstructed by $\gamma\pi^+\pi^-$ and $\eta\pi^+\pi^-$ in subplots (a) and (b), respectively. The line shape of the PHSP process is determined by a Monte Carlo simulation in Ref.~\cite{BESIII:2018zbm}.}
    \label{fig:1mm}
\end{figure*}

\begin{figure*}[t]
    \centering
    \includegraphics[width=0.8\linewidth]{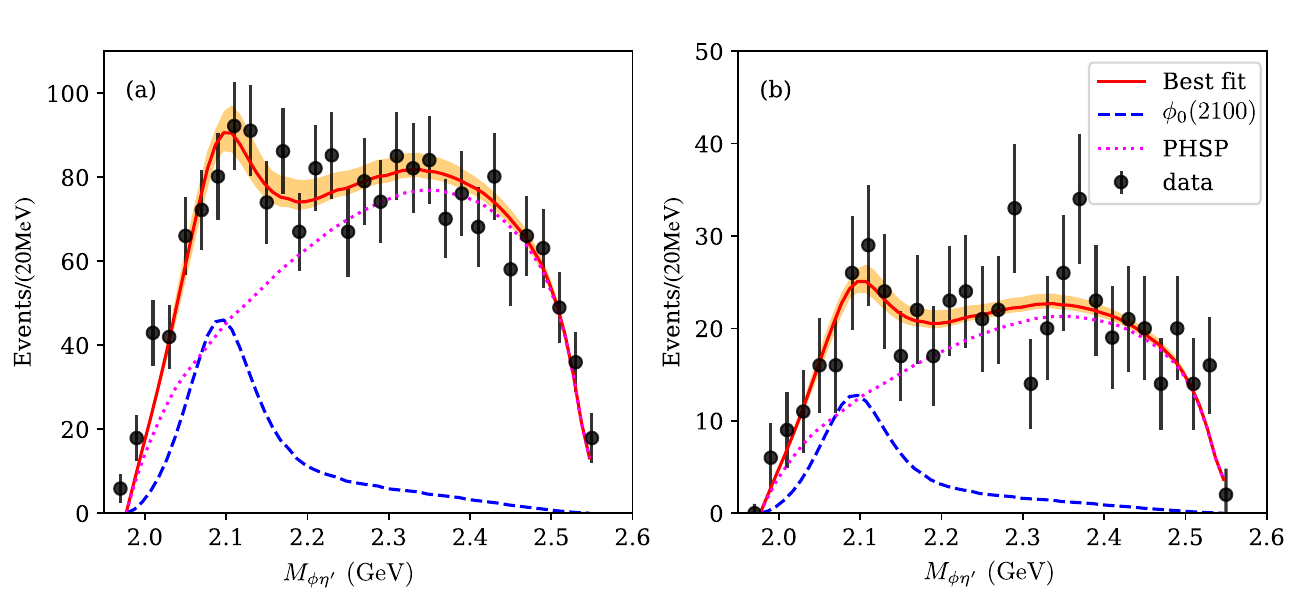}
    \caption{The best fit of the $J/\psi\to\phi\eta'\eta$ data~\cite{BESIII:2018zbm} with $0^{--}$ $K^*\bar K_1$ rescattering only in the $\phi\eta'$ channel ($P_b=0$). See the caption of Fig.~\ref{fig:1mm}.}
    \label{fig:0mm}
\end{figure*}

Only the invariant mass distribution of $\phi\eta'$ is published in Ref.~\cite{BESIII:2018zbm} where the enhancement near 2.1~GeV was described by a BW resonance. Since the $\phi\eta$ invariant mass distribution was not reported, we will first try to fit the data in Ref.~\cite{BESIII:2018zbm} by considering only the $K^*\bar K_1$ rescattering in the $\phi\eta'$ channel, i.e., $P_b$ is fixed to 0.
The reconstruction of the $\eta'$ in Ref.~\cite{BESIII:2018zbm} involves two modes, where the $\eta'$ is reconstructed by $\gamma\pi^+\pi^-$ and $\eta\pi^+\pi^-$, respectively. These two data sets are simultaneously fitted using the same differential decay width, as shown in Eq.~\eqref{eq:dGamma}. However, to take into account the efficiency difference between these two modes, a normalization factor $\beta$ is introduced. In total, there are 5 free parameters: $P_a$, $P_c$, $\alpha$, $\beta$ and $C^{(1)}$ or $C^{(0)}$. The parameters obtained from the best fits are listed in Table~\ref{tab:pars} with $P_b=0$ fixed and the fitting results are shown in Figs.~\ref{fig:1mm} and~\ref{fig:0mm}. We can see that both assumptions, $J^{PC}=1^{--}$ and $0^{--}$, provide a satisfactory description of the  data. With $C^{(1)}$ and $C^{(0)}$ from the best fits, the pole positions of the $K^*\bar K_1$ molecules are determined to be $(2079-68i)$~MeV for $1^{--}$, denoted by $\phi(2100)$ and $(2091-61i)$~MeV for $0^{--}$, denoted by $\phi_0(2100)$.

\begin{table*}[t]
	\centering
	\caption{\label{tab:pars} The parameters from the best fits together with the pole positions, $E_b$, relative to the $K^*\bar K_1$ threshold at 2145~MeV.
    }
    \begin{ruledtabular}
    \begin{tabular}{|c|ccccccccc|}

				$ J^{PC}  $ & $\Lambda$ (GeV) & $P_a\ (10^{3})$ & $P_b\ (10^{3})$& $P_c\ (10^{3})$ & $\alpha$ & $\beta$ & $C^{(1/0)}$& $\chi^2/$d.o.f. &$E_B$ (MeV) \\
				\hline
				\multirow{2}{*}{$1^{--}$}&  $1$ (fixed) & $-1.6\pm 0.9 $& 0 (fixed)& $37.0\pm2.9$ &$1.44\pm0.06$ & $0.28\pm0.02$ & $-2.5\pm4.5$& $38.1/(60-5)$ & $-66-68i$\\
                &  $1$ (fixed) & 0 (fixed) & $23.6\pm5.4$ & $29.1\pm 11.2$ &$1.39\pm0.06$ & $0.28\pm0.02$ & $-0.7\pm 4.6$& $35.1/(60-5)$ & $-60-68i$\\
                \hline
                \multirow{2}{*}{$0^{--}$}&  $1$ (fixed) & $-1.4\pm 1.1 $ & 0 (fixed) & $33.4\pm2.5$ &$1.48\pm0.07$ & $0.28\pm0.02$ & $-22\pm6$& $37.6/(60-5)$ & $-54-61i$\\
                &  $1$ (fixed) & 0 (fixed) & $21.0\pm4.6$ &$30.3\pm 10.3$ &$1.41\pm0.06$ & $0.28\pm0.02$ & $-20\pm4$& $32.1/(60-5)$ & $-48-61i$\\
   \end{tabular}
   \end{ruledtabular}
\end{table*}


The quantum numbers of the introduced resonance were analyzed in Ref.~\cite{BESIII:2018zbm} by examining the $\eta$ polar angular distribution. If the quantum numbers $J^{P}$ of the introduced resonance in the $\phi\eta'$ channel are $1^{+}$, $1^{-}$, or $0^{-}$, the $\eta$ polar angular distribution is proportional to $1$, $1+\cos^2\theta$, or $\sin^2\theta$, respectively. It is found in Ref.~\cite{BESIII:2018zbm} that both the assumptions of $J^P=1^+$ and $1^{-}$ for the resonance in the $\phi\eta'$ channel can describe the data, with the former being more preferred. However, the assumption of $J^P=0^-$ was excluded as it seemed to deviate significantly from the data in the analysis of Ref.~\cite{BESIII:2018zbm}. 

\begin{figure}
    \centering
    \includegraphics[width=0.9\linewidth]{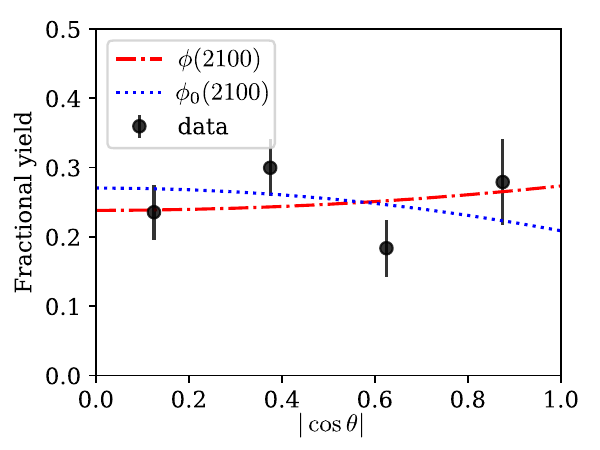}
    \caption{The $\eta$ polar angular distribution in $J/\psi\to\phi\eta'\eta$. The data are taken from Ref.~\cite{BESIII:2018zbm} and the lines are the predictions from the best fits shown in Figs.~\ref{fig:1mm} and~\ref{fig:0mm}.}
    \label{fig:ang}
\end{figure}

It is important to note that the contribution of the PHSP process, in both Ref.~\cite{BESIII:2018zbm} and our fit result, are much larger than that of the introduced resonance. However, the $\eta$ polar angular distribution from the PHSP process is not settled based on the published data in Ref.~\cite{BESIII:2018zbm}, and it is not necessarily the same as that of the resonance. The authors in Ref.~\cite{BESIII:2018zbm} did not consider the contribution of the PHSP process to the $\eta$ polar angular distribution. Here we assume that the $\eta$ polar angular distribution from the PHSP process is flat as a consequence of the $S$-wave nature of the background term in Eq.~\eqref{eq:dGamma}, the total $\eta$ polar angular distribution can be predicted as follows:
\begin{equation}
    \frac{\mathrm{d}\Gamma}{\mathrm{d}\cos\theta}\propto \frac14\left\{ \begin{array}{cc}
       \left(\tilde\alpha_1 + \frac34(1-\tilde\alpha_1)(1+\cos^2\theta)\right)  & \text{for }1^{--} \\
        \left(\tilde\alpha_0 + \frac34(1-\tilde\alpha_0)(1-\cos^2\theta)\right) & \text{for }0^{--}
    \end{array}\right.,\label{eq:polarang}
\end{equation}
where $\tilde\alpha_1=0.815$ and $\tilde\alpha_0=0.835$ represent the fraction of the PHSP process obtained in our $1^{--}$ and $0^{--}$ fits, respectively. The comparison between the predictions in Eq.~\eqref{eq:polarang} and the data are shown in Fig.~\ref{fig:ang}. From Fig.~\ref{fig:ang}  it is evident that both $1^{--}$ and $0^{--}$ assumptions provide a satisfactory description of the data. Consequently, we cannot definitely conclude whether the resonance signal is from the $\phi(2100)$, $\phi_0(2100)$, or  that both manifest in the $J/\psi\to\phi\eta'\eta$ decay. To address this, we propose to conduct an analysis of this process using the complete set of $J/\psi$ events recorded by the BESIII detector~\cite{BESIII:2021cxx}, which is one order of magnitude larger than the sample size utilized in the previous study~\cite{BESIII:2018zbm}. The difference of the two curves in Fig.~\ref{fig:ang} may be disentangled with the full dataset. Furthermore, performing a partial wave analysis on the polar angular distribution of the $\eta$, as well as the helicity angular distribution of $\phi\eta'$, one may be able to ascertain the presence of the $1^{--}$ and $0^{--}$ $K^*\bar K_1$ bound states. The information obtained from the $\phi\eta$ channel is also of great value, as the $K^*\bar K_1$ molecules can also decay into $\phi\eta$.

\subsection{$K^*\bar K_1$ in both $\phi\eta'$ and $\phi\eta$ channels}
\begin{figure*}[t]
    \centering
    \includegraphics[width=0.8\linewidth]{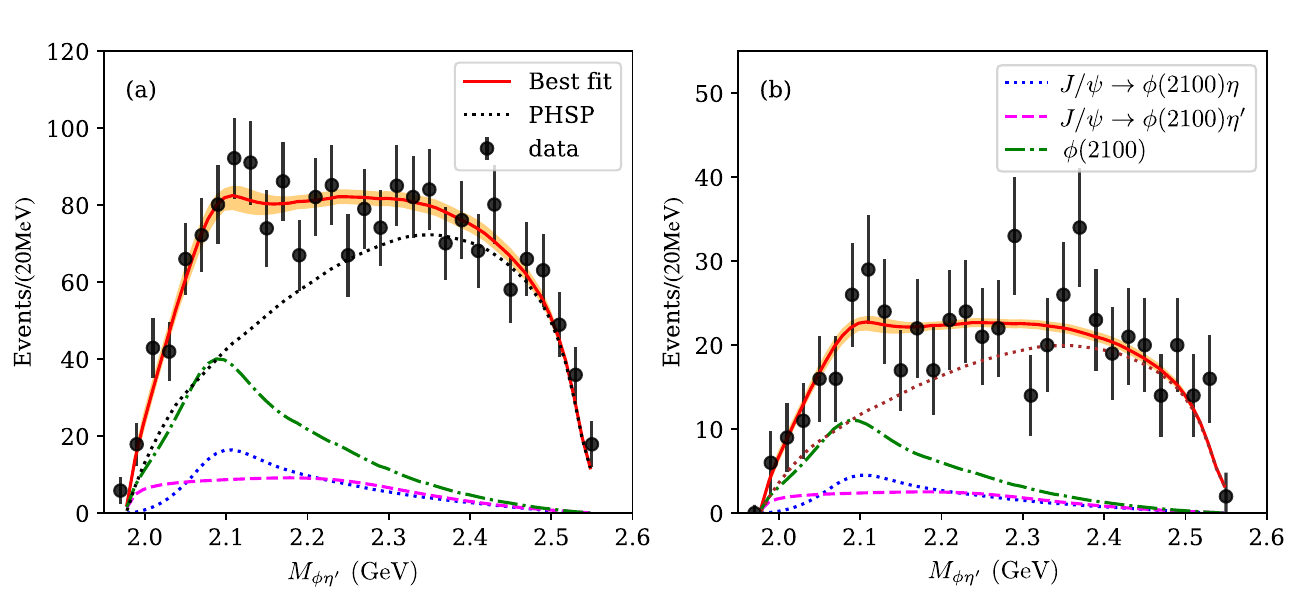}
    \caption{The best fit of the $J/\psi\to\phi\eta'\eta$ data~\cite{BESIII:2018zbm} with $1^{--}$ $K^*\bar K_1$ rescattering in both $\phi\eta'$ and $\phi\eta$ channels. The green dot-dashed line represents the full contribution of $\phi(2100)$ while the blue dotted and magenta dashed lines represent the individual contributions of $\phi(2100)$ in $J/\psi\to\phi(2100)\eta, \phi(2100)\to\phi\eta'$ and $J/\psi\to\phi(2100)\eta', \phi(2100)\to\phi\eta$, respectively. See the caption of Fig.~\ref{fig:1mm}.}
    \label{fig:1mmp3}
\end{figure*}

\begin{figure*}[t]
    \centering
    \includegraphics[width=0.8\linewidth]{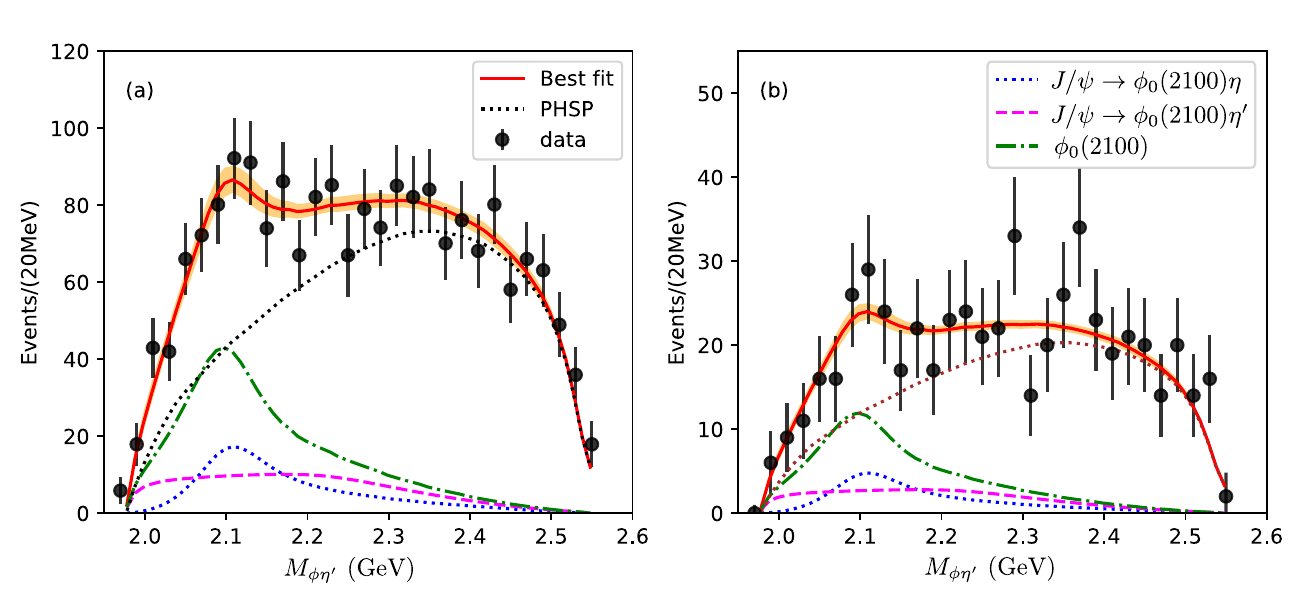}
    \caption{The best fit of the $J/\psi\to\phi\eta'\eta$ data~\cite{BESIII:2018zbm} with $0^{--}$ $K^*\bar K_1$ rescattering in both $\phi\eta'$ and $\phi\eta$ channels. The green dot-dashed line represents the full contribution of $\phi_0(2100)$ while the blue dotted and magenta dashed lines represent the individual contributions of $\phi_0(2100)$ in $J/\psi\to\phi_0(2100)\eta, \phi_0(2100)\to\phi\eta'$ and $J/\psi\to\phi_0(2100)\eta', \phi_0(2100)\to\phi\eta$, respectively. See the caption of Fig.~\ref{fig:1mm}.}
    \label{fig:0mmp3}
\end{figure*}

In this subsection we try to include the contribution of $K^*\bar K_1$ from the $\phi\eta$ channel by letting $P_b$ free. As discussed before and confirmed by the fits in the previous subsection, the  $P_a$ term is of higher order. To reduce the number of parameters, we fix $P_a=0$ in the following calculation. We still have $5$ free parameters in total, $P_b$, $P_c$, $\alpha$, $\beta$ and $C^{(1)}$ or $C^{(0)}$. The parameters from the best fit are listed in Table~\ref{tab:pars} with $P_a=0$ fixed. The fitting results are shown in Figs.~\ref{fig:1mmp3} and~\ref{fig:0mmp3}. The resulting pole positions of the $\phi(2100)$ and the $\phi_0(2100)$ hardly change.

\begin{figure*}[t]
    \centering
    \includegraphics[width=0.8\linewidth]{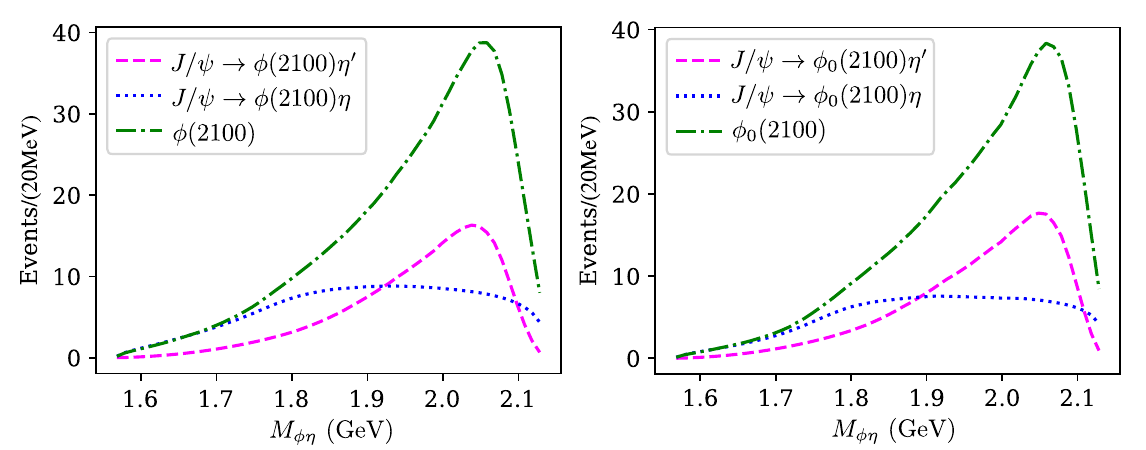}
    \caption{Predicted $\phi\eta$ invariant mass distribution in the $J/\psi\to\phi\eta'\eta$ decay  contributed from the $\phi(2100)$ (left) or $\phi_0(2100)$ (right) with $K^*\bar K_1$ rescattering in both $\phi\eta'$ and $\phi\eta$ channels. See the caption of Fig.~\ref{fig:1mmp3} for the meaning of each line.}
    \label{fig:phieta}
\end{figure*}

The $\phi\eta$ invariant mass distributions of $J/\psi\to\phi\eta'\eta$ decay are also predicted by the following expression,
\begin{align}
\frac{\mathrm{d}\Gamma_{J/\psi\to\phi\eta'\eta}}{\mathrm{d}M_{\phi\eta}}=\int{\rm d}M_{\phi\eta'}^2\frac{2M_{\phi\eta}}{256\pi^3m^3_{J/\psi}}|T_{J/\psi\to\phi\eta'\eta}|^2,\label{eq:dGammaphieta}
\end{align}
where only the contribution of the $\phi(2100)$ or the $\phi_0(2100)$ is included since the lineshape of the PHSP contribution in the $\phi\eta$ invariant mass distribution is not available from the published data in Ref.~\cite{BESIII:2018zbm}. The predicted $\phi\eta$ invariant mass distribution is shown in Fig.~\ref{fig:phieta}, and one sees a peak near 2.1~GeV. 
In fact, from the Dalitz plot reported by the BESIII Collaboration~\cite{BESIII:2018zbm}, there seems a accumulation of events at $m_{\phi\eta}\simeq 2.1$~GeV.


\section{Summary}\label{sec:summary}
The interaction between $K^*$ and $\bar K_1$ has been investigated in the OBE model, where  $t$-channel vector meson and pseudoscalar meson exchanges are taken into account. {There are two parameters in the potential to absorb the cutoff dependence of physical observables  and they can be determined by experimental data.} Additionally, the $u$-channel $\pi$ exchange, which plays a crucial role in the decay width of the $K^*\bar K_1$ molecule, is also considered, where the three body effects of $K^*\bar K^*\pi$ are carefully examined. 

In order to investigate the cause of the observed enhancement near 2.1~GeV in the $\phi\eta'$ final states in the $J/\psi\to\phi\eta'\eta$ process~\cite{BESIII:2018zbm}, we conduct a fit analysis of the invariant mass distribution of $\phi\eta'$. The inclusion of the $K^*\bar K_1$ channel in the analysis yields satisfactory results, as both the $\phi(2100)$ and the $\phi_0(2100)$ are able to adequately describe the data. However, it is difficult to determine whether one or both of these states contribute to the $J/\psi\to\phi\eta'\eta$ decay, even when considering the $\eta$ polar angular distribution. It is worth noting that the $K^*\bar K_1$ bound states are also capable of decaying into $\phi\eta$. Therefore, valuable insights into the $K^*\bar K_1$ bound states can be obtained by analyzing the invariant mass distribution of $\phi\eta$, which is predicted in Fig.~\ref{fig:phieta}. The Dalitz plots in Ref.~\cite{BESIII:2018zbm} reveal an accumulation of data within the range of $[4,4.5]\ \rm GeV^2$ in the $\phi\eta$ final states. Consequently, we propose conducting a study on the $J/\psi\to\phi\eta'\eta$ decay using the entire dataset of $J/\psi$ events collected by BESIII~\cite{BESIII:2021cxx}, which is approximately eight times larger than the dataset used in Ref.~\cite{BESIII:2018zbm}. By performing a partial wave analysis of the polar and helicity angular distributions, one may be able to disentangle the contribution of $\phi(2100)$ and $\phi_0(2100)$ to the $J/\psi\to\phi\eta'\eta$ decay. Furthermore, other decays of $J/\psi$ into $\eta K\bar K$, $\eta K^*\bar K^*$,  $\eta K\bar K^*$ and $\phi\eta\eta$ can also be explored to study the resonance(s) around 2.1~GeV. While the $\phi(2100)$ should contribute to all these processes, the $\phi_0(2100)$ can only couple to the last two.

\begin{acknowledgments}
X.-K. Dong would like to thank Xiao-Yu Li for valuable discussions about the experimental details. This work is supported in part by the Chinese Academy of Sciences under Grants No.~XDB34030000 and No. YSBR-101; by the National Natural Science Foundation of China (NSFC) under Grants No. 12125507, No. 12361141819 and No. 12047503; by the National Key R\&D Program of China under Grant No. 2023YFA1606703; by the Deutsche Forschungsgemeinschaft (DFG, German Research Foundation) and the NSFC through the funds provided to the Sino-German Collaborative Research Center TRR110 ``Symmetries and the Emergence of Structure in QCD'' (DFG Project ID 196253076 - TRR 110, NSFC Grant No. 12070131001); and by the CAS President's International Fellowship Initiative (PIFI) (Grant No. 2018DM0034).

\end{acknowledgments}

\begin{appendix}
\section{Details of the $S$-wave projection}\label{app:SwavePro}
When solving the LSE for the $S$-wave scattering amplitude, we should first project the potential onto $S$-wave by
\begin{align}
    V_S(k',k)=\frac12\int_{-1}^{+1} {\rm d}z\ V(q)~\label{eq:Sproj}
\end{align}
with $\bm q^2=k^2+k'^2-2kk'z$ and $z$ the angle between the incoming and outgoing particles. In the LSE, the on-shell potential $V_S(k_0,k_0)$, the half-on-shell potential $V_S(k_0,k_i)$ and the off-shell potential $V_S(k_i,k_j)$ are all needed, where $k_0$ is the on-shell momentum of $K^*$ in the c.m. of $K^*\bar K_1$ determined by the equation 
\begin{align}
    E- {k_0}^{2} /\left(2 \mu\right)+i \Gamma(E; {k_0})/2=0,
\end{align}
$k_i$ is the momentum lies on the integral path, $[0,+\infty)$, of Eq.~\eqref{eq:LSE0}. Note that $k_0$ is complex in general while $k_i$ is always a real positive value.

The scattering amplitude is an analytical function of complex $E$ except for possible cuts and poles. It is the amplitude on the physical axis, i.e., the real axis on the physical Riemann sheet, that affects the lineshapes observed in experiments. Such amplitude is obtained by taking the integral path of $z$ in Eq.~\eqref{eq:Sproj} to be $[-1,1]$ along the real $z$ axis. In practice, when the pole of the integrand, say $z_0$, is close to the integral path $[-1,1]$, we deform the integral path away from the pole for better numerical performance. On the other hand, when searching for poles on the complex energy plane, $E$ should move from the physical axis to the pole position so that this pole has direct influences on the amplitude on the physical axis. In this process, $z_0$ may also move and possibly cross the integral path $[-1,1]$, which will result in a discontinuity, namely a cut, of the amplitude. To cross this cut continuously, the integral path should be deformed accordingly to avoid the cross with the trajectory of $z_0$. This is the main logic to obtain the amplitude that is connected to the physical axis directly. 

We will show how to choose the integral path for the $S$-wave projection of the potential from $t$-channel and $u$-channel meson exchanges in the following. Note that when calculating the amplitude on the physical axis, all the off-shell, half-on-shell and on-shell potentials are needed while when searching for poles on the physical Riemann sheet,\footnote{On the physical Riemann sheet, poles can only appear on the real axis below the lowest threshold for stable constituent particles, and it will move to the lower half energy plane if the constituent particles have a finite width.} only the off-shell potentials are relevant. Therefore, we can pay attention just to the trajectory of $z_0$ when varying $E$ for the off-shell potential.

\subsection{$t$-channel meson exchange}
The potential from the $t$-channel meson exchange takes the form of 
\begin{align}
    V^t(\bm q)\propto \frac{1}{k^2+k^{\prime2}-2kk'z+m^2-i\epsilon},
\end{align}
and the pole position of the integrand reads
\begin{align}
    z_0=\frac{k^2+k^{\prime2}+m^2-i\epsilon}{2kk'}.
\end{align}
For the off-shell potential where $k,k'\ge0$, $z_0$ is independent of $E$ and lies beyond the integral path $[-1,1]$. For the on-shell or half-on-shell potential, $z_0$ has an imaginary part due to the finite width of $K^*$ and $K_1$, and hence $z_0$ is also far from the integral path $[-1,1]$. Therefore, the integral path of the $t$ channel meson exchange needs no deformation.

\subsection{$u$-channel $\pi$ exchange}
The analytical structure is much more complicated for the $u$-channel $\pi$ exchange. In the LSE, the propagator of the exchanged $\pi$ in Eq.~\eqref{eq:Vpiu} should be rewritten as
\begin{align}
    \frac{1}{q^2-m_\pi^2+i \epsilon} \rightarrow \frac{1}{2 W\!\left(m_\pi, \boldsymbol{q}\right)}\left(\frac{1}{d_1}+\frac{1}{d_2}\right),
\end{align}
where
\begin{align}
    d_i=\sqrt{s}-W\!\left(m_\pi, \boldsymbol{q}\right)-W\!\left(m_i, \boldsymbol{k}\right)-W\!\left(m_i, \boldsymbol{k}^{\prime}\right),
\end{align}
with $\sqrt s$ the total energy, $m_1=m_{K^*}-i\Gamma_{K^*}/2$, $m_2=m_{K_1}-i\Gamma_{K_1}/2$ and $W(m,\boldsymbol{q})=\sqrt{\boldsymbol{q}^2+m^2}$. 

\begin{figure}
    \centering
    \includegraphics[width=\linewidth]{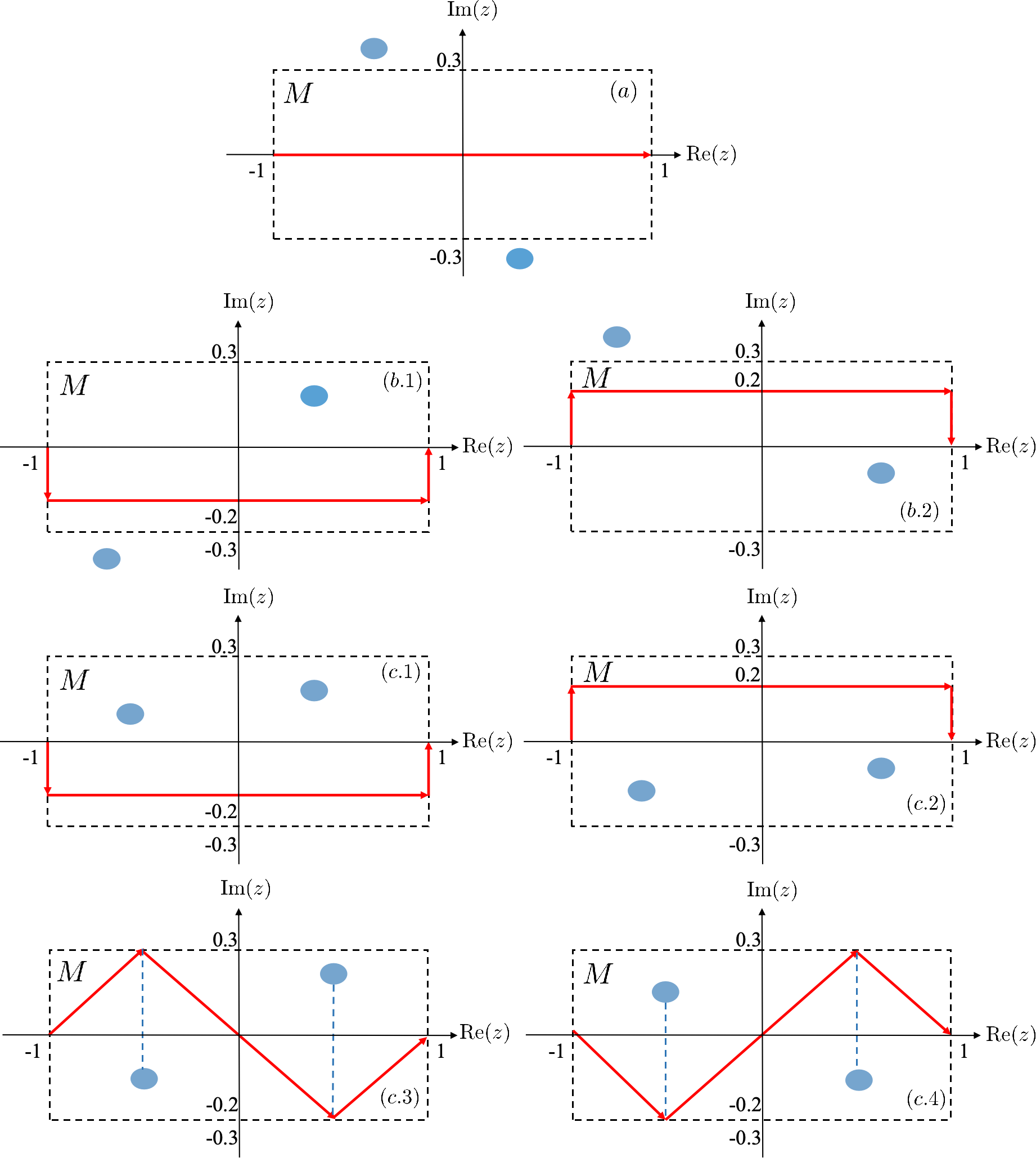}
    \caption{Possible pole positions, $z_{0d}$ and $z_{0\pi}$, and the corresponding integral path to obtain continuous results on the physical axis.}
    \label{fig:pathintegral}
\end{figure}

When performing the $S$-wave projection for the on-shell and half-on-shell potentials, the poles of the integrand may come from the roots of $d_1$ and $W\!\left(m_\pi, \boldsymbol{q}\right)$, denoted by $z_{0d}$ and $z_{0\pi}$, respectively. We should first determine the pole positions and then deform the integral path accordingly to make the poles not so close to the integral path, as shown in Fig.~\ref{fig:pathintegral}. Precisely, we define a region $M$ where $-1<$Re[$z$]$<1$ and $-0.3<$Im[$z$]$<0.3$, and then choose the integral path properly  by determining if $z_{0d}$ and $z_{0\pi}$ belong to $M$. Specific situations can be divided into several cases:
\begin{itemize}
    \item[(a)] $z_{0d}  \notin M$ \& $z_{0\pi} \notin M$, such as (a) in Fig.~\ref{fig:pathintegral};
    \item[(b)] ($z_{0d}  \in M$ \& $z_{0\pi} \notin M$) $| |$ ($z_{0d}  \notin M$ \& $z_{0\pi} \in M$):
    \begin{itemize}
        \item[(1)] Im[$z_M$]$>0$, such as (b.1) in Fig.~\ref{fig:pathintegral},
        \item[(2)] Im[$z_M$]$<0$, such as (b.2) in Fig.~\ref{fig:pathintegral},
    \end{itemize}
    where the one contained in $M$ is denoted as $z_M$;
    \item[(c)] $z_{0d}  \in M$ \& $z_{0\pi} \in M$:
    \begin{itemize}
        \item[(1)] Im[$z_{0d}$]$>0$ \& Im[$z_{0\pi}$]$>0$, such as (c.1) in Fig.~\ref{fig:pathintegral},
        \item[(2)] Im[$z_{0d}$]$<0$ \& Im[$z_{0\pi}$]$<0$, such as (c.2) in Fig.~\ref{fig:pathintegral},
        \item[(3)] Im[$z_{0d}$]$>0$ \& Im[$z_{0\pi}$]$<0$  or  Im[$z_{0d}$]$<0$ \& Im[$z_{0\pi}$]$>0$, such as (c.3) or (c.4) in Fig.~\ref{fig:pathintegral}, respectively.
    \end{itemize}
\end{itemize}

When performing the $S$-wave projection for the off-shell potentials, we can easily obtain the cuts on complex energy plane,
\begin{align}
    \sqrt{s}\in [\sqrt{s_-},\sqrt{s_+}],
\end{align}
with $\sqrt{s_{\pm}}=\sqrt{m_\pi^2+(k\pm k')^2}+W\!\left(m_1, \boldsymbol{k}\right)+W\!\left(m_1, \boldsymbol{k}^{\prime}\right)$. 
These cuts are line segments on the lower half energy plane. Therefore, the amplitude at energy $\sqrt s$ is calculated with the integral path $z\in[-1,1]$ along the real axis if ${\rm Im}(\sqrt s)>{\rm Im}(\sqrt{s_\pm})$ or ${\rm Re}(\sqrt s)\notin {\rm Re}([\sqrt{s_-},\sqrt{s_+}])$. Otherwise, we need to cross the cut continuously by adding the residue of the pole, {multiplied by $2\pi i$,} to the result obtained by using the integral path $z\in[-1,1]$ along real axis. 

\end{appendix}

\bibliography{ref}

\end{document}